\documentclass[a4paper,11pt,DIV=12]{scrartcl}
\pdfoutput=1

\usepackage[ttscale=0.9]{libertine}
\usepackage{mathrsfs}
\usepackage[intlimits]{amsmath}
\usepackage{amssymb}
\usepackage{slashed}
\usepackage[titletoc,title]{appendix}
\usepackage[affil-it]{authblk}
\usepackage[numbers,sort&compress]{natbib}
\usepackage{graphicx}
\usepackage[normalem]{ulem}
\usepackage{layouts}
\usepackage[protrusion=true,expansion,kerning=true,tracking=true,final]{microtype}
\usepackage[%backref,
	        colorlinks=true,
	        linkcolor=hblue,
	        citecolor=hgreen,
	        filecolor=hblue,
	        urlcolor=hred
	        ]{hyperref}
\usepackage[dvipsnames,x11names]{xcolor}
\definecolor{hgreen}{rgb}{0,.3,0}
\definecolor{hred}{rgb}{.3,0,0}
\definecolor{orange}{rgb}{1,0.5,0}
\definecolor{hblue}{rgb}{0,0,.3}
\definecolor{LightGray}{gray}{0.95}
\definecolor{gray}{gray}{0.6}

\usepackage[T1,LY1]{fontenc}

\makeatletter
\DeclareOldFontCommand{\rm}{\normalfont\rmfamily}{\mathrm}
\DeclareOldFontCommand{\sf}{\normalfont\sffamily}{\mathsf}
\DeclareOldFontCommand{\tt}{\normalfont\ttfamily}{\mathtt}
\DeclareOldFontCommand{\bf}{\normalfont\bfseries}{\mathbf}
\DeclareOldFontCommand{\it}{\normalfont\itshape}{\mathit}
\DeclareOldFontCommand{\sl}{\normalfont\slshape}{\@nomath\sl}
\DeclareOldFontCommand{\sc}{\normalfont\scshape}{\@nomath\sc}

\usepackage{scrlayer-scrpage}
\allowdisplaybreaks
\setcapindent{1em}

\setkomafont{captionlabel}{\bfseries}
\setkomafont{caption}{\itshape}
\setkomafont{titlehead}{\normalsize\sffamily}
\addtokomafont{title}{\boldmath}
\addtokomafont{section}{\boldmath}

\KOMAoptions{headinclude=false,
             footinclude=false,
             twoside=false,
             parskip=false,
             draft=false,
             abstract=true,
             numbers=noenddot,
             DIV=12}

\definecolor{Blu}{rgb}{0.,0.,1.}
\definecolor{Red}{rgb}{1.,0.,0.}
\definecolor{Darkgreen}{rgb}{0,0.5,0.}
\definecolor{Purple}{rgb}{0.5,0.,0.5}

\newcommand{\mumu}{(\mu^+\mu^-)_{\ell=0}}

\begin{document}

\titlehead{\hfill DO-TH 22/24}
\title{Impact of indirect CP violation on Br$(K_S \to \mu^+\mu^-)_{\ell=0}$}
\date{\today}
\renewcommand\Authands{, }
\author[a]{Joachim~Brod%
        \thanks{\texttt{joachim.brod@uc.edu}}}
\author[b]{Emmanuel~Stamou%
        \thanks{\texttt{emmanuel.stamou@tu-dortmund.de}}}

\affil[a]{{\large Department of Physics, University of Cincinnati, Cincinnati, OH 45221, USA}}
\affil[b]{{\large Fakult\"at f\"ur Physik, TU Dortmund, D-44221 Dortmund, Germany}}

\maketitle

\begin{abstract}
%\normalsize
  The decay $K_S \to (\mu^+\mu^-)_{\ell=0}$, with the final muon pair
  in an angular-momentum zero state, is a sensitive probe of
  short-distance physics. It has recently been shown how to extract
  this branching ratio from neutral kaon decay data. We point out that
  the impact of indirect CP violation on the standard-model prediction
  of this mode, while nominally of order $|\epsilon_K| \sim 10^{-3}$,
  is enhanced by a large amplitude ratio and leads to a shift of the
  branching ratio Br$(K_S \to \mu^+\mu^-)_{\ell=0}$ by a few percent,
  depending on the size of a relative phase that can be extracted from
  data. We also update the standard-model prediction of the
  short-distance contribution.
  %  In total we find
  % Br$(K_S \to \mu^+\mu^-)_{\ell=0}^{\text{SM}} = 
  % 1.70\,(02)_\text{QCD/QED} (01)_{f_K} (06)_\text{ICPV} (19)_\text{param.} \times 10^{-13}$.
\end{abstract}

%\pdfbookmark[1]{Table of Contents}{tableofcontents}
\setcounter{page}{1}
%\tableofcontents
%    \the\textwidth\\
%    \printinunitsof{cm}\prntlen{\textwidth}\\
%    \printinunitsof{in}\prntlen{\textwidth}

\section{Introduction\label{sec:introduction}}

CP violation in neutral kaon decays is a very sensitive probe of
short-distance dynamics, both within the standard model of particle
physics (SM) and beyond. A prime example is the rare decay
$K_L \to \pi^0 \nu \bar \nu$. It proceeds almost exclusively via CP
violation in interference between mixing and
decay~\cite{Littenberg:1989ix, Grossman:1997sk} and is theoretically
exceptionally clean~\cite{Mescia:2007kn, Brod:2010hi}.  A dedicated
experimental program to measure its branching ratio is
underway~\cite{KOTO:2020prk}.

The amplitude for the decay of neutral kaons into a charged dimuon
final state, $K\to \mu^+\mu^-$, also has a short-distance component;
however, it has been believed for a long time that these modes are
less sensitive to high-energy dynamics due to very large long-distance
(LD) contributions that are hard to control
theoretically~\cite{Isidori:2003ts}.  However, it has been pointed out
in Ref.~\cite{DAmbrosio:2017klp} that the interference term between
the $K_L$ and $K_S$ components in the time-dependent decay rate of
neutral kaons into dileptons receives a large CP-violating
contribution, primarily sensitive to short-distance (SD) dynamics.

The dimuon final state resulting from the disintegration of a $K_L$
can have angular-momentum $\ell=0$ or $\ell=1$; we denote these final
states by $(\mu^+ \mu^-)_{\ell}$. The time-dependent decay rate into
either final state of an initial neutral kaon that was tagged as a
state $K$ at time $t=0$ is given by~\cite{Dery:2021mct,
  Workman:2022ynf}
\begin{equation}\label{eq:decay}
\begin{split}
  \frac{1}{{\mathcal N}_\ell}  \frac{d\Gamma(K(t) \to (\mu^+ \mu^-)_{\ell})}{dt} = 
  C_L^\ell e^{-\Gamma_L t} + C_S^\ell e^{-\Gamma_S t}
  + 2 \big[ C_\text{sin}^\ell \sin(\Delta M t) + C_\text{cos}^\ell \cos(\Delta M t) \big] e^{-\Gamma t}
\end{split}
\end{equation}
where $\Gamma_S$ and $\Gamma_L$ are the $K_S$ and $K_L$ total decay
rates, $\Gamma = (\Gamma_S + \Gamma_L)/2$, and $\Delta M = M_L - M_S$,
with $M_S$ and $M_L$ the $K_S$ and $K_L$ masses. ${\mathcal N}_\ell$
is a conventional, final-state dependent normalization factor. The
parameters $C_L^\ell$, $C_S^\ell$, $C_\text{cos}^\ell$, and
$C_\text{sin}^\ell$ depend on the initial and final states and can be
experimentally determined. In practice, the $\ell=0$ and $\ell=1$
final states are experimentally indistinguishable. Since there is no
interference between these two final states, it follows that only the
incoherent sum of the two decay rates can be measured. More
explicitly, only the combinations
$C_i \equiv C_i^{\ell=0} + \beta_\mu^2 C_i^{\ell=1}$, where
$i=L,S,\text{cos}, \text{sin}$, can be measured (the relative factor
$\beta_\mu^2$ accounts for the phase-space difference of the $\ell=0$
and $\ell=1$ final states, see Eq.~\eqref{eq:Gamma} below).

Recently, it has been shown in Ref.~\cite{Dery:2021mct} how to use
interference data from $K_{L/S} \to \mu^+ \mu^-$ to extract the
angular-momentum $\ell = 0$ part of the branching fraction of $K_S$
into a pair of muons, Br$(K_S \to \mu^+ \mu^-)_{\ell = 0}$.  The main
result of Ref.~\cite{Dery:2021mct} is the relation 
\begin{equation}\label{eq:DGGS}
  \text{Br}(K_S \to  \mu^+ \mu^-)_{\ell = 0} = \text{Br}(K_L \to \mu^+ \mu^-)
  \times \frac{\tau_S}{\tau_L}
  \times \bigg( \frac{C_\text{int}}{C_L} \bigg)^2 \,,
\end{equation}
where $\tau_S(\tau_L)$ are the $K_S(K_L)$ lifetimes, and
$C_\text{int} \equiv (C_\text{cos}^2 + C_\text{sin}^2)^{1/2}$. 
Eq.~\eqref{eq:DGGS} is valid for a pure $K^0$ or
$\bar K^0$ beam; see Ref.~\cite{Dery:2021mct} for a discussion of more
general cases. {\itshape Within the SM}, the branching ratio
Br$(K_S \to \mu^+ \mu^-)_{\ell = 0}$ is fully dominated by SD
contributions and can be calculated perturbatively; the remaining
theoretical and parametric uncertainties are small.  A precise
measurement of Br$(K_S \to \mu^+ \mu^-)_{\ell = 0}$ via
Eq.~\eqref{eq:DGGS} would, therefore, constitute a sensitive test of
the SM.

In this work, we will critically assess the assumptions made in
deriving Eq.~\eqref{eq:DGGS}. In particular, we will scrutinize the
assumption that there is no indirect CP violation in neutral kaon
mixing, made in Ref.~\cite{Dery:2021mct}. We will show in
Sec.~\ref{sec:rates} that the relation Eq.~\eqref{eq:DGGS} remains
true in the presence of indirect CP violation. As a consequence, the
$K_S \to (\mu^+ \mu^-)_{\ell = 0}$ branching ratio extracted using
Eq.~\eqref{eq:DGGS} includes the effect of indirect CP violation, and
this effect has to be taken into account in its SM prediction
to perform a meaningful comparison. 
In fact, we will show that the effect, which is naively of order $\epsilon_K \sim 10^{-3}$, can
be surprisingly large.

Sec.~\ref{sec:br} contains the main result of this work. We update the
numerical value of the SD contribution to the
$K_S \to (\mu^+ \mu^-)_{\ell = 0}$ branching ratio, based on the known
next-to-next-leading-order QCD and next-to-leading-order electroweak
corrections~\cite{Bobeth:2013tba, Hermann:2013kca, Bobeth:2013uxa}.
Moreover, we point out that the branching ratio for
$K_S \to (\mu^+ \mu^-)_{\ell = 0}$ can receive a sizeable LD
correction due to indirect CP violation in the neutral kaon
system. While this additional term is naively suppressed by a factor
of
$|\epsilon_K| = (2.228 \pm 0.011) \times
10^{-3}$~\cite{Workman:2022ynf}, it is enhanced by a large ratio of
amplitudes, leading to a correction of up to $\pm 3.7\%$.  We show how
this LD contribution can, in principle, be fully determined from data
and estimate its {\itshape maximal} size using the measurement of
Br$(K_L\to\mu^+\mu^-)$.

To understand the size of this correction, it may be instructive to
recall that an analogous mechanism is at play in the rare decay
$K_L \to \pi^0 \nu \bar \nu$. It was shown in
Ref.~\cite{Buchalla:1996fp} that the contribution of indirect CP
violation to this CP-violating decay is proportional to $|\epsilon_K|$
times the {\em real} (CP-conserving) part of the effective
Lagrangian, involving charm-quark contributions in addition to the
leading top-quark contribution, as well as different CKM input
parameters. The net result is an enhancement by roughly a factor of
three over the naive estimate, leading to a percent correction of the
branching ratio~\cite{Buchalla:1996fp, Brod:2010hi}.
For the decay $K_S \to (\mu^+ \mu^-)_{\ell=0}$, a similar mechanism is
a play. In this case, however, the real part of the amplitude
receives, in addition to a contribution from the real part of the
short-distance Lagrangian, a large CP-conserving contribution of a LD
two-photon intermediate state. This leads to a much larger correction
than naively expected from the smallness of $\epsilon_K$.
Our conclusions are contained in Sec.~\ref{sec:conclusions}.

\section{Decay rate and CP structure\label{sec:rates}}

In general, a neutral kaon $K_j = K_S, K_L$ decaying into a dimuon
final state produces the lepton pair either in the CP-odd
angular-momentum $\ell = 0$ state or in the CP-even angular-momentum
$\ell = 1$ state.  The corresponding total decay rate is given
by~\cite{Cirigliano:2011ny}
\begin{equation}\label{eq:Gamma}
  \Gamma(K_j \to \mu^+ \mu^-)
  = \frac{M_{K_j}}{8\pi} \beta_\mu (\beta_\mu^2 |A_1^j|^2 + |A_0^j|^2) \,,
\end{equation}
with $\beta_\mu \equiv (1 - 4m_\mu^2/M_{K_j}^2)^{1/2}$. This defines
our normalisation for the angular-momentum amplitudes
$A_\ell^j \equiv \langle (\mu^+ \mu^-)_{\ell} | K_j \rangle$. In other
words, the term proportional to $|A_0^j|^2$ corresponds to the lepton
pair being produced in the CP-odd $\ell = 0$ state, and the one
proportional to $|A_1^j|^2$ to it being produced in the CP-even
$\ell = 1$ state.

The flavour and mass eigenstates of the neutral kaons are related by
\begin{equation}\label{eq:KSKL}
  |K_S\rangle = p | K^0\rangle + q |\bar K^0\rangle\,, \qquad
  |K_L\rangle = p | K^0\rangle - q |\bar K^0\rangle\,.
\end{equation}
Using this decomposition, we can express the experimental parameters
of the time-dependent decay rate in Eq.~\eqref{eq:decay} in terms of
the $K_L/K_S$ amplitudes.\footnote{
For kaon mixing, we find it more transparent to express the $C$'s 
in terms of the amplitudes of the mass eigenstates, rather 
than the flavor eigenstates as in Refs.~\cite{Workman:2022ynf,Dery:2021mct}.
} 
In the PDG conventions \cite{Workman:2022ynf},
this implies for $K(t=0) = K^0$ (``$K^0$ beam'')
\begin{align}
  C_L^\ell &= \frac{1}{2|p|^2} |A_\ell^L|^2\,,&
  C_{\rm sin}^\ell & = \frac{1}{2|p|^2}\text{Im} \big\{\big(A_\ell^S\big)^* A_\ell^L \big\} \,,\nonumber
  \\ 
  C_S^\ell &= \frac{1}{2|p|^2}|A_\ell^S|^2 \,,&
  C_{\rm cos}^\ell &= \frac{1}{2|p|^2}\text{Re} \big\{\big(A_\ell^S\big)^* A_\ell^L \big\} \,,
  \label{eq:CK0}
\end{align}
and analogously for $K(t=0) = \bar K^0$ (``$\bar K^0$ beam'') 
\begin{align}
  C_L^\ell &= \frac{1}{2|q|^2} |A_\ell^L|^2\,,&
  C_{\rm sin}^\ell & = -\frac{1}{2|q|^2}\text{Im} \big\{\big(A_\ell^S\big)^* A_\ell^L \big\} \,,\nonumber
  \\ 
  C_S^\ell &= \frac{1}{2|q|^2}|A_\ell^S|^2 \,,&
  C_{\rm cos}^\ell &= -\frac{1}{2|q|^2}\text{Re} \big\{\big(A_\ell^S\big)^* A_\ell^L \big\} \,.
  \label{eq:CK0bar}
\end{align}
Eqs.~\eqref{eq:CK0} and \eqref{eq:CK0bar} agree with the corresponding
equations in Ref.~\cite{Dery:2021mct}, which are written in terms of
the flavour-basis amplitudes. Note that, according to
Eq.~\eqref{eq:Gamma}, the normalisation factors for the $\ell=0$ and
$\ell=1$ final states are related by 
${\mathcal N}_1 = \beta_\mu^2 {\mathcal N}_0 = 0.82 {\mathcal N}_0$.
Note that the global $1/|p|^2$ and $1/|q|^2$ factors can 
be absorbed into the normalisation factor ${\mathcal N}_{\ell}$. 
We only keep them to conform with the PDG conventions \cite{Workman:2022ynf}.

With these definitions, it is now straightforward to derive
Eq.~\eqref{eq:DGGS} {\em under the sole assumption that $A_1^L = 0$.}
We have
\begin{equation}\label{eq:so:easy:to:derive}
\begin{split}
  \frac{\text{Br}(K_S \to  \mu^+ \mu^-)_{\ell = 0}}
  {\text{Br}(K_L \to \mu^+ \mu^-)_{\phantom{\ell=0}}}
& = \frac{|A_0^S|^2}{|A_0^L|^2} \times \frac{\tau_L}{\tau_S}
= \frac{|A_0^S|^2|A_0^L|^2}{|A_0^L|^4} \times \frac{\tau_L}{\tau_S}
= \left(\frac{C_\text{int}}{C_L}\right)^2 \times \frac{\tau_L}{\tau_S} \,,
\end{split}
\end{equation}
where we used that $C_L = C_L^0$ and $C_\text{int} = C_\text{int}^0$
if $A^L_1 = 0$. It is clear from this derivation that
Eq.~\eqref{eq:DGGS} is true also in the presence of indirect CP
violation, since we use the amplitudes for the exact weak mass
eigenstates. The presence of indirect CP violation is
not something that can be switched off by choice --- it is experimental fact,
and it will necessarily affect the experimental determination of
$\text{Br}(K_S \to\mu^+ \mu^-)_{\ell = 0}$ based on
Eq.~\eqref{eq:DGGS}. It follows that indirect CP violation should be
taken into account in the SM prediction of this branching ratio. This
is the main purpose of this work.

\bigskip

As a preparation for the remainder of this work
%, and in order to better understand the condition $A_1^L = 0$,
we now briefly discuss the CP structure of the decays
$K_j \to \mu^+ \mu^-$. First, we consider the case that indirect CP
violation is absent in $K \to \mu^+\mu^-$ (i.e. $\epsilon_K = 0$). For
clarity, we will then explicitly use the even and odd CP eigenstates
\begin{equation}\label{eq:K1K2}
  |K_1 \rangle = \tfrac{1}{\sqrt{2}} \big( |K^0 \rangle - |\bar K^0 \rangle \big) \,, \qquad
  |K_2 \rangle = \tfrac{1}{\sqrt{2}} \big( |K^0 \rangle + |\bar K^0 \rangle \big) \,,
\end{equation}
where we adopted phase conventions such that
${\rm CP}|K^0 \rangle = - |\bar K^0 \rangle$,
${\rm CP}|\bar K^0 \rangle = - |K^0 \rangle$. We have
$|K_S \rangle \to |K_1 \rangle$ and $|K_L \rangle \to |K_2 \rangle$ in
the limit $\epsilon_K \to 0$.  In the SM, the genuine SD contributions
are induced by the effective Lagrangian\footnote{ We use a
  normalization of the effective Lagrangian conforming with the
  convention in Ref.~\cite{Bobeth:2013uxa} to transparently include
  perturbative corrections to $Y_t$ (see section~\ref{sec:sm}).}
\begin{equation}\label{eq:Leff}
  {\mathcal L}_\text{eff,~SD}^{|\Delta S|=1}
= \frac{2G_F^2 M_W^2}{\pi^2}
  \big( 
     V_{ts}^* V_{td}^{\phantom{*}} Y_t
   + V_{cs}^* V_{cd}^{\phantom{*}} Y_\text{NL} \big)
  Q_\mu + \text{h.c.} \,.
\end{equation}
Here, $Y_\text{NL}$ and $Y_t$ are functions of
$x_t \equiv m_t^2/M_W^2$ and $x_c \equiv m_c^2/M_W^2$, respectively,
with $m_t$ and $m_c$ the top- and charm-quark masses, $M_W$ the
$W$-boson mass, and the local operator is defined as
\begin{equation}
  Q_\mu = (\bar s_L \gamma^\nu d_L)(\bar \mu_L \gamma_\nu \mu_L) \,.
\end{equation}

This effective Lagrangian originates from electroweak box and penguin
diagrams~\cite{Buchalla:1995vs} and generates only the
$A_{\ell = 0}^j$ amplitudes.  Hence, in the limit of vanishing
indirect CP violation its contribution is CP-conserving for
$K_2 \to \mu^+ \mu^-$ and CP-violating for $K_1 \to \mu^+
\mu^-$. However, within the SM, the muon pair can also be produced via
a two-photon intermediate state originating from operators of the full
$|\Delta S|=1$ Lagrangian other than $Q_\mu$.  Contrary to the $Q_\mu$
contribution, the two-photon intermediate state generates both the
${\ell = 0}$ and the ${\ell = 1}$ amplitudes.  The two-photon
contribution is completely LD dominated and nearly CP
conserving~\cite{Isidori:2003ts}; hence, the ${\ell = 1}$ amplitude
can be neglected for $K_2 \to \mu^+ \mu^-$, but not for
$K_1 \to \mu^+ \mu^-$~\cite{Isidori:2003ts}.

An important comment is in order.  As we have shown, the only
assumption in deriving Eq.~\eqref{eq:DGGS} is the vanishing of the
amplitude $A_L^1$.  Since $A_S^1$ is non-zero, indirect CP violation
is expected to induce a small but non-zero contribution to the
amplitude $A_L^1$, of order $\epsilon_K \times A_S^1$.  (This
contribution has to be added to any amplitude
$K_2 \to (\mu^+ \mu^-)_{\ell=1}$ that may be present due to direct CP
violation.)  We therefore expect $A_1^L$ to be non-zero, with an
absolute size that is small but hard to quantify. With $A_1^L$
nonzero, we can no longer identify the experimentally determined
parameters $C_L$ and $C_\text{int}$ with $C_L^0$ and $C_\text{int}^0$,
respectively; hence, Eq.~\eqref{eq:DGGS} is no longer exactly
valid. Since the numerical value of $A^L_1$ is not well known, it is
hard to make quantitative statements, but based solely on the
contribution to $A_1^L$ arising from indirect CP violation, the
relation in Eq.~\eqref{eq:DGGS} will receive a correction at the
percent level.\footnote{If we loosen the assumption $A_1^L=0$, the
  relation in Eq.~\eqref{eq:DGGS} receives corrections because in this
  case $C_L\neq C_L^0$ and $C_{\text{int}}\neq C^0_{\text{int}}$ (see
  the derivation in Eq.~\eqref{eq:so:easy:to:derive}).  In the limit
  of $|A^L_1|\ll |A^L_0|$, the leading corrections multiplying the right
  side of Eq.~\eqref{eq:DGGS} are
    \begin{equation*}
      r = 1 - 2 \beta_\mu^2 \frac{|A^L_1|}{|A^L_0|} \frac{|A^S_1|}{|A^S_0|}\cos(\phi_0\!-\!\phi_1)
      + \beta_\mu^4 \frac{|A^L_1|^2}{|A^L_0|^2} \frac{|A^S_1|^2}{|A^S_0|^2}
        \big(4 \cos^2(\phi_0\!-\!\phi_1) - 1\big)
      +{\cal O}\left(  {|A^L_1|}^3/{|A^L_0|}^3\right)\,,
    \end{equation*}
    with the strong phases
    $\phi_\ell \equiv \arg\big\{\big(A_\ell^S\big)^* A_\ell^L\big\}$.
    To estimate the size of the corrections, we use
    $|A_0^S| = 2.64 \times 10^{-13}$ and
    $|A_0^L| = 2.22 \times 10^{-12}$ which we derive below, and the
    estimate in Ref.~\cite{DAmbrosio:2017klp} for the long-distance
    contribution to the branching ratio for $K_L \to \mu^+ \mu^-$ to
    obtain $|A_1^S| = 1.58 \times 10^{-12}$, with an unspecified
    uncertainty.  For the correction to be below a percent we must
    have $|A_1^L|\lesssim 2\times 10^{-15}$ for
    $\cos(\phi_0\!-\!\phi_1)\sim {\cal O}(1)$ and
    $|A_1^L| \lesssim 5\times 10^{-14}$ if
    $\cos(\phi_0\!-\!\phi_1)\sim 0$. This can be compared to a naive
    estimate for $A_1^L$, namely,
    $A_1^L = A_1^L|_{\epsilon_K=0} + \epsilon_K A_1^S$.  Using
    $A_1^L|_{\epsilon_K=0} = 0$, this gives
    $|A_1^L| \sim 3.52 \times 10^{-15}$. We see that corrections to
    Eq.~\eqref{eq:DGGS} of the order of one percent are expected.}
Further study of this issue seems worthwhile; this is,
however, beyond the scope of this work.

\section{Br$(K_S \to \mu^+\mu^-)_{\ell=0}$ in the Standard Model\label{sec:br}}
In this section, we will show that the amplitude $A_0^S$ receives two
contributions: the first involves only the imaginary part of the
effective Lagrangian, and the second, proportional to $\epsilon_K$,
involves the real part of the effective Lagrangian. Throughout, we
neglect all effects that are of higher powers in $\epsilon_K$.

To obtain this decomposition we recall that, in the presence of
indirect CP violation, the neutral kaon mass eigenstates are related
to the flavor eigenstates via Eq.~\eqref{eq:KSKL}.
%\begin{equation}\label{eq:KSKL}
%  |K_S\rangle = p | K^0\rangle + q |\bar K^0\rangle\,, \qquad
%  |K_L\rangle = p | K^0\rangle - q |\bar K^0\rangle\,,
%\end{equation}
For the following argument, it will be convenient to switch to the
``traditional'' notation\footnote{We neglected the tiny correction
  factor $1/\sqrt{1+|\bar{\epsilon}|^2}$ in the coefficients.}
$p = (1 + \bar\epsilon)/\sqrt{2}$,
$q = - (1 - \bar\epsilon)/\sqrt{2}$, with the approximation
$\bar\epsilon \approx \epsilon_K$~\cite{Buchalla:1996fp}. This gives
\begin{equation}\label{eq:KSKL:trad}
\begin{split}
  | K_S \rangle  
  = \frac{1+\epsilon_K}{\sqrt{2}} | K^0 \rangle - \frac{1-\epsilon_K}{\sqrt{2}} | \bar K^0 \rangle \,,\qquad
  | K_L \rangle
  = \frac{1+\epsilon_K}{\sqrt{2}} | K^0 \rangle + \frac{1-\epsilon_K}{\sqrt{2}} | \bar K^0 \rangle \,,
\end{split}
\end{equation}
and we can write (note that
${\mathcal L}_\text{eff}^{\Delta S = -1} = \big( {\mathcal
  L}_\text{eff}^{\Delta S = 1} \big)^\dagger$)
\begin{equation}\label{eq:0:KS}
\begin{split}
A_0^S \equiv \langle \mumu | {\mathcal L}_\text{eff}^{|\Delta S| = 1} | K_S \rangle
 = &+\frac{1+\epsilon_K}{\sqrt{2}} \langle \mumu | {\mathcal L}_\text{eff}^{\Delta S = 1} | K^0 \rangle\\
   &-\frac{1-\epsilon_K}{\sqrt{2}} \langle \mumu | {\mathcal L}_\text{eff}^{\Delta S = -1}| \bar K^0 \rangle \,.
\end{split}
\end{equation}
Here, we explicitly display the flavor-changing weak interactions,
induced by higher dimension operators, as these are the only terms
that potentially involve a weak phase; i.e.,
${\mathcal L}_\text{eff}^{|\Delta S| = 1}$ denotes the full
$|\Delta S| = 1$ effective Lagrangian, not just the $Q_\mu$
operator. All effects of the strong and electromagnetic interactions
are understood to be taken into account implicitly in the matrix
elements. Using the CP invariance of QCD and QED, together with the
well-known transformation properties of currents and states under CP
(see, e.g., Ref.~\cite{Anikeev:2001rk}), it is straightforward to show
that
\begin{equation}\label{eq:relation}
  \langle \mumu | {\mathcal L}_\text{eff}^{\Delta S = -1} | \bar K^0 \rangle
= \langle \mumu | \big( {\mathcal L}_\text{eff}^{\Delta S = 1} \big)^* | K^0 \rangle \,.
\end{equation}
Note that the complex conjugation acts only on the Wilson coefficients
in the Lagrangian. Combining Eqs.~\eqref{eq:0:KS}
and~\eqref{eq:relation}, we obtain\\[-2em]
\begin{equation}\label{eq:KS:K0}
\begin{split}
A_0^S \equiv   \langle \mumu | {\mathcal L}_\text{eff}^{|\Delta S| = 1} | K_S \rangle
  = \overbrace{
    \sqrt{2} i          \langle \mumu | {\rm Im}\big({\mathcal L}_\text{eff}^{\Delta S = 1}\big) | K^0 \rangle
  }^{=A^S_0\big\vert_{\epsilon_K=0}~\text{(see section~\ref{sec:sm})}}&\\
    + 
    \underbrace{
    \sqrt{2} \epsilon_K \langle \mumu | {\rm Re}\big({\mathcal L}_\text{eff}^{\Delta S = 1} \big)| K^0 \rangle
  }_{\text{(see section~\ref{sec:ld})}}\ &\,.
\end{split}
\end{equation}
We see that, in the first term, only the imaginary part of the Wilson
coefficients in the effective Lagrangian contributes. To the extent
that we assume that the LD contributions are CP conserving, this term
arises only from the operator $Q_\mu$ and can be calculated using
perturbation theory, see Sec.~\ref{sec:sm}. The second term,
proportional to $\epsilon_K$, is dominated by LD contributions that
are hard to calculate. We will show in Sec.~\ref{sec:ld} that they can
be estimated from data.

\subsection{Perturbative CP-violating contribution from ${\rm Im}({\mathcal L}_{\rm eff}^{|\Delta S| = 1})$\label{sec:sm}}

The contribution to Br$(K_S\to\mu^+\mu^-)_{\ell=0}$ proportional to
$\big\vert A^S_0\vert_{\epsilon_K=0}\big\vert^2$ from the first term
in Eq.~\eqref{eq:KS:K0} involves the imaginary part of the
$|\Delta S|=1$ Wilson coefficients.  It is thus proportional to the
short-distance top-quark contribution of the effective Lagrangian in
Eq.~\eqref{eq:Leff} proportional to
${\rm Im}(V_{ts}^*V_{td}) \times Y_t = A^2\lambda^5\bar \eta\times
Y_t$.  In what follows, we denote this contribution to the branching
ratio by $\text{Br}(K_S \to \mu^+ \mu^-)_{\ell = 0}^\text{pert.}$.

The $Y_t$ function receives perturbative QCD and EW corrections.  With
respect to QCD, there is no renormalization-group (RG) evolution for
the $Q_\mu$ operator, thus the QCD corrections for $Y_t$ are the same
as for the $B_q\to\mu^+\mu^-$ decays.  They have been computed with
next-to-next-to-leading accuracy in Ref.~\cite{Hermann:2013kca}; the
residual scale uncertainty is below $0.1\%$ on $Y_t$.  Here, we simply
update the numerical value for the top quark. The
next-to-leading-order EW corrections have been computed for the
$B_q\to\mu^+\mu^-$ decays in Ref.~\cite{Bobeth:2013tba}.  In this
case, there is a mixed QCD$\times$QED RG evolution that involves
operator mixing.  To consistently include the EW corrections we extend
the evolution down to $2$\,GeV (see Ref.~\cite{Huber:2005ig} and
Appendix B in Ref.~\cite{Bobeth:2013tba}); the residual scheme and
scale uncertainty is $0.5\%$ on $Y_t$.  Including both
next-to-next-to-leading-order QCD and next-to-leading-order EW
corrections we find for $\mu_{\text{low}} = 2$\,GeV:
\begin{equation}
  Y_t = 0.931 \pm 0.001\big\vert_\text{QCD} \pm 0.005\big\vert_\text{EW}\,,
  \label{eq:Yt}
\end{equation}
where the uncertainties corresponding to residual scale and scheme uncertainties.
All input is taken from Ref.~\cite{Workman:2022ynf}, in particular we use for the
top-quark mass $M_t^{\text{pole}} = 172.5(7)$\,GeV from the cross-section measurements.
The $W$-boson mass is not a primary input, we calculate it as a
function of the $Z$-boson mass, the Higgs-boson mass, and
the strong and the electromagnetic coupling constants $\alpha_s$ and
$\alpha$, respectively (see Ref.~\cite{Awramik:2003rn}).

The CP-violating contribution to the decay $K_S \to (\mu^+ \mu^-)_{\ell=0}$ from
the imaginary part of the effective Lagrangian in Eq.~\eqref{eq:Leff} then reads
\cite{Isidori:2003ts, Dery:2021mct}
\begin{equation}
  \text{Br}(K_S \to \mu^+ \mu^-)_{\ell = 0}^\text{pert.}
= \frac{\beta_\mu \tau_S}{16\pi M_K} A^4 \lambda^{10} \bar\eta^2
  \bigg|
        \frac{2G_F^2 M_W^2}{\pi^2} f_K M_K m_\mu Y_t 
  \bigg|^2 \,.
\end{equation}
Using Eq.~\eqref{eq:Yt}, $f_K = 155.7(3)\,$MeV~\cite{Aoki:2021kgd},
and all remaining input parameters from Ref.~\cite{Workman:2022ynf}
(apart from the $W$-boson mass that is not an independent input
parameter -- see above) we find
\begin{equation}
  \text{Br}(K_S \to \mu^+ \mu^-)_{\ell = 0}^\text{pert.}
= 1.70\, (02)_\text{QCD/EW} (01)_{f_K} (19)_\text{param.} \times 10^{-13}\,.
\end{equation}
See the discussion at the end of Sec.~\ref{sec:ld} for the error
budget of the parametric uncertainties. Using the normalisation as in
Eq.~\eqref{eq:Gamma}, this branching ratio corresponds to
$\big|A_0^S|_{\epsilon_K = 0} \big| = 2.64 \times 10^{-13}$.

\subsection{Contribution proportional to $\epsilon_K$ and ${\rm Re}({\mathcal L}_{\rm eff}^{|\Delta S| = 1})$}\label{sec:ld}

After having calculated the perturbative contribution to the
$K_S \to \mumu$ decay rate, it remains to estimate the correction
proportional to $\epsilon_K$. We recall that the $K_L \to \mu^+ \mu^-$
decay rate is fully dominated by the amplitude $A_0^L$.
To relate this amplitude to the term in Eq.~\eqref{eq:KS:K0} proportional to
$\epsilon_K$, we use Eq.~\eqref{eq:relation} to rewrite
\begin{equation}
  \begin{split}
  2 \langle \mumu | \text{Re}\big({\mathcal L}_\text{eff}^{\Delta S = 1}\big) | K^0 \rangle
  & = \langle \mumu | {\mathcal L}_\text{eff}^{\Delta S = 1} | K^0 \rangle
    + \langle \mumu | \big( {\mathcal L}_\text{eff}^{\Delta S = 1} \big)^* | K^0 \rangle \\
  & = \langle \mumu | {\mathcal L}_\text{eff}^{\Delta S = 1} | K^0 \rangle
    + \langle \mumu | {\mathcal L}_\text{eff}^{\Delta S = -1} | \bar K^0 \rangle \\
    & = \sqrt{2} \underbrace{\langle \mumu | {\mathcal L}_\text{eff}^{|\Delta S| = 1} | K_L \rangle}_{\equiv A_0^L}
      + {\cal O}(\epsilon_K)\,.
  \end{split}
\end{equation}
Combining this with Eq.~\eqref{eq:KS:K0} we have
\begin{equation}
\begin{split}
  A_0^S = A_0^S\big|_{\epsilon_K=0} + \epsilon_K A_0^L  + {\mathcal O}(\epsilon_K^2) \,.
\end{split}
\end{equation}

For the decay rate, we need the absolute value squared of the
amplitude. Keeping terms up to first order in $\epsilon_K$, we find
\begin{equation}
\begin{split}
  \big| A_0^S \big|^2 
  = \Big| A_0^S \big|_{\epsilon_K=0}  \Big|^2
  + 2 \,{\rm Re}
      \Big\{ \Big( A_0^S \big|_{\epsilon_K=0} \Big)^* A_0^L \epsilon_K\Big\} 
      +{\mathcal O}(\epsilon_K^2)\,.
\end{split}
\end{equation}
To linear order in $\epsilon_K$, we can approximate
$A_0^S \big|_{\epsilon_K=0}$ by just $A_0^S$ in the term proportional
to $\epsilon_K$. Defining the relative phase
$\phi_0 = \arg\big\{\big(A_0^S\big)^* A_0^L\big\}$ as in
Ref.~\cite{Dery:2021mct}, we have
\begin{equation}
\begin{split}
\text{Re}
\Big\{ \Big( A_0^S \Big)^* A_0^L \epsilon_K \Big\}
=  \frac{|\epsilon_K|}{\sqrt{2}} |A_0^S| \, |A_0^L| (\cos\phi_0 - \sin\phi_0) \,,
\end{split}
\end{equation}
where we used~\cite{Workman:2022ynf}
\begin{equation}
  \epsilon_K \approx \frac{1+i}{\sqrt{2}} |\epsilon_K| \,.
  \label{eq:eKapprox}
\end{equation}
Therefore, the branching ratio including the effects of indirect CP
violation is obtained via
\begin{equation}
\begin{split}
  \text{Br}(K_S \to \mu^+ \mu^-)_{\ell=0}
= & \, \text{Br}(K_S \to \mu^+ \mu^-)_{\ell=0}^\text{pert.} \\
  & \times
  \bigg( 1 + \sqrt{2} |\epsilon_K|\frac{|A_0^L|}{|A_0^S|}
  (\cos\phi_0 - \sin\phi_0)  \bigg) \,,
\end{split}
\end{equation}
up to tiny corrections of order $\epsilon_K^2$. Even without further
knowledge about the size of the phase~$\phi_0$ we can estimate the
maximal size of the correction to the branching ratio, by observing
that $|\cos\phi_0 - \sin\phi_0| \leq \sqrt{2}$ and extracting
$|A_0^L|$ from data ($|A_0^S| \simeq 2.64 \times 10^{-13}$ has been
calculated in the previous section). Using the fact that the
$K_L \to \mu^+ \mu^-$ decay is completely dominated by the $\ell = 0$
amplitude, the value of $|A_0^L|$ can be extracted from the measured
branching ratio
Br$(K_L \to \mu^+ \mu^-) = (6.84 \pm 0.11) \times
10^{-9}$~\cite{Workman:2022ynf}. Using Eq.~\eqref{eq:Gamma} and the
$K_L$ lifetime
$\tau_L = 5.116(21) \times 10^{-8}\,s$~\cite{Workman:2022ynf}, we find
$|A_0^L| \simeq 2.22(2) \times 10^{-12}$.  Together with
$|\epsilon_K| = (2.228 \pm 0.011) \times
10^{-3}$~\cite{Workman:2022ynf}, this leads to a maximal correction
factor of $( 1 \pm 0.037)$ such that
\begin{equation}
  \text{Br}(K_S \to \mu^+ \mu^-)_{\ell=0}
  =  \, \text{Br}(K_S \to \mu^+ \mu^-)_{\ell=0}^\text{pert.} \times r_\text{ICPV} \,,
\end{equation}
where
\begin{equation}
  r_\text{ICPV} = 1 + 0.037 \, \, \frac{\cos\phi_0 - \sin\phi_0}{\sqrt{2}} \,.
\end{equation}
Accordingly, without any knowledge of the phase, the SM prediction of
$\text{Br}(K_S \to \mu^+ \mu^-)_{\ell=0}$ is afflicted with an
uncertainty of up to $\pm 3.7\%$ that was previously not accounted
for. 

Interestingly, it has been shown recently how to extract precise
information on the relative phase $\phi_0$ from $K_L \to \mu^+ \mu^-$
and $K_L \to \gamma \gamma$ data~\cite{Dery:2022yqc}. It was found
that $\cos^2\phi_0 = 0.96 \pm 0.03$, where the error includes both
experimental and theoretical uncertainties, see
Ref.~\cite{Dery:2022yqc} for details. This leads to four allowed
values for the relative phase $\phi_0$, with two corresponding to a
negative sign of $\cos\phi_0$, and another two corresponding to a
positive sign. The negative sign is preferred from theoretical
arguments, as discussed in detail in Refs.~\cite{Isidori:2003ts,
  Dery:2022yqc}.
g
The resulting branching ratios are shown in Tab.~\ref{tab:br}. For the
CKM input we have used the current SM fit from
Ref.~\cite{Workman:2022ynf} giving
$\lambda = 0.22500(67)$,
$A=0.826^{+0.018}_{-0.015}$,
$\bar\rho = 0.159(10)$,
$\bar\eta = 0.348(10)$.
We see that the non-parametric uncertainty, arising from unknown
higher-order perturbative corrections and the kaon decay constant, is
of the order of $1\%$, similar as for the $K_L \to \pi^0 \nu \bar \nu$
mode. The dominant (relative) parametric uncertainties arise from the
CKM input parameters and are $8.7\%$ from $A$, $5.7\%$ from
$\bar\eta$, $3.0\%$ from $\lambda$, as well as $1.3\%$ from the top
quark mass; the remaining parametric uncertainties (including those
related to the relative phase $\phi_0$) are at the permil level and
thus negligible.

We note in this context that in Ref.~\cite{Buras:2021nns} the ratio
$R_\text{SL} = \text{Br}(K_S \to \mu^+ \mu^-)_{\ell=0}/\text{Br}(K_L
\to \pi^0 \nu \bar\nu)$ has been suggested as a null-test of the SM
that is free of CKM input parameters.\footnote{The dependence of this
  ratio on $\lambda^2$, as it appears in Ref.~\cite{Buras:2021nns}, is
  spurious in the sense that it assumes the specific parameterization
  of the hadronic matrix elements for the $K_{\ell 3}$ decays employed
  in the fit in Ref.~\cite{Mescia:2007kn}.} The effect of indirect CP
violation calculated here amounts to a small 
correction to $R_\text{SL}$ of the order of a few percent,
in which, however, the CKM parameters do not cancel.

\begin{table}
\centering
\begin{tabular}{cccc}
  $\cos(\phi_0)$ & $\sin(\phi_0)$ & $r_{\text{ICPV}}$ & $\text{Br}(K_S \to \mu^+ \mu^-)_{\ell = 0}$ \\ \hline\hline\\[-0.8em]
  $\phantom{+}0.98$ & $\phantom{+}0.20$ & $1.021$ & $1.74(2)(19) \times 10^{-13}$ \\[0.2em] 
  $\phantom{+}0.98$ & $-0.20$ & $1.031$ & $1.75(2)(19) \times 10^{-13}$ \\ [0.2em]
  $-0.98$ & $\phantom{+}0.20$ & $0.969$ & $1.65(2)(18) \times 10^{-13}$ \\ [0.2em]
  $-0.98$ & $-0.20$ & $0.979$ & $1.67(2)(18) \times 10^{-13}$ \\[0.25em]\hline
\end{tabular}
\caption{Predictions for the $K_S \to (\mu^+ \mu^-)_{\ell = 0}$
  branching ratio in dependence on the different values for the strong
  phase $\phi_0$. The first uncertainty comprises the missing
  higher-order perturbative corrections as well as the uncertainty in
  $f_K$, while the second uncertainty is parametric (dominated by the
  uncertainty in the CKM input parameters).}
\label{tab:br}
\end{table}

\section{Discussion and conclusions\label{sec:conclusions}}

The decay $K_S \to (\mu^+ \mu^-)_{\ell=0}$ is
almost purely CP violating and dominated, in the SM and many of its
extensions, by short-distance contributions that can be calculated
perturbatively with high precision. It has been demonstrated in
Refs.~\cite{DAmbrosio:2017klp, Dery:2021mct} that this branching
ratio can, in principle, be extracted from neutral kaon interference
data, thus adding a new precision test of the SM in the kaon sector.

In this work, we revisited the SM prediction of
$\text{Br}(K_S \to \mu^+ \mu^-)_{\ell=0}$. To the extent that indirect
CP violation in the neutral kaon sector is neglected, the only
relevant hadronic input parameter is the kaon decay constant, which is
know with permil accuracy. The SD contribution is comprised by the
loop function $Y_t$~\cite{Buchalla:1995vs}; results for the NNLO
QCD~\cite{Hermann:2013kca} and NLO electroweak~\cite{Bobeth:2013tba}
corrections are available in the literature and have been combined
here into a state-of-the art SM prediction.

Upon close inspection, it turns out that Eq.~\eqref{eq:DGGS}, relating
the searched-for branching ratio to measurable quantities, comprises
the effects of indirect CP violation. It is, therefore, mandatory to
include these effects also in the SM prediction. Maybe somewhat
surprisingly, the impact of indirect CP violation on the branching
ratio is enhanced over the naive estimate $\epsilon_K \sim 10^{-3}$ by
about an order of magnitude, due to non-perturbative effects of the
strong interactions that are hard to calculate. Estimating the size of
the correction with the help of available data, we found that this can
shift the SM prediction by up to $\pm 3.7\%$; an effect that was
previously unaccounted for. Using a recent estimate of the relative
phase $\phi_0$ that was determined up to a four-fold ambiguity, we
find that the correction to the branching ratio is either $\pm 2\%$ or
$\pm 3\%$ (see Tab.~\ref{tab:br}), with the negative values preferred
from theoretical arguments.

Finally, we briefly summarize the assumptions that enter at various
stages of the analysis (see Ref.~\cite{Dery:2021mct} for details).  We
start with an assumption that we {\em do not} need to make: as shown
in this work, CP violation in neutral kaon mixing can consistently be
taken into account.  We made, however, the following assumptions:
{\itshape (i)} the only source of CP violation is in the $\ell = 0$
amplitude; and {\itshape (ii)} the long-distance physics is CP
conserving.  Both assumptions are satisfied to very good approximation
in the SM.  Assumption {\itshape (i)} is also valid in all SM
extensions in which the leading operator is vectorial.

\section*{Acknowledgments}

We thank Andrzej Buras, Giancarlo D'Ambrosio, Avital Dery, Yuval
Grossman, Christopher Smith, and Jure Zupan for discussions.  JB
thanks the Physics Department at the Weizmann Institute for Science
for hospitality, and acknowledges support in part by DoE grant
DE-SC0011784.

\addcontentsline{toc}{section}{References}
\bibliographystyle{JHEP}
\bibliography{references}

\end{document}